\documentstyle[prd,aps,epsf]{revtex}
\begin{document}

\input epsf \renewcommand{\topfraction}{0.8}
\twocolumn[\hsize\textwidth\columnwidth\hsize\csname
@twocolumnfalse\endcsname

\title{Tunnelling with Bianchi IX Instantons.}
\author{P. M. Saffin}
\address{DAMTP,
            Silver Street,
            Cambridge.\\ 
            E-mail: p.m.saffin@damtp.cam.ac.uk
}
\date{\today}
\maketitle
\begin{abstract}
Within the context of finding the initial conditions of the
universe we consider gravitational instantons falling into
the Bianchi IX classification. That is, a Euclidean four-manifold
with a metric that satisfies Einstein's equations with an induced 
metric on S$^3$ submanifolds that is homogeneous but
anisotropic. As well as finding regular solutions to the field
equations with a tunnelling scalar field, we also look at the case
of singular instantons with a view to applying the results to
generic potentials.
The study is in agreement with the prejudice that instantons with 
higher symmetry have a lower Euclidean action, even when we consider
the singular class of solutions. It is also found that the Euclidean
action can diverge for simple potentials, showing that the Hawking
Turok instanton had finite action owing to its symmetry. 
\end{abstract}

\vskip2pc]


\section{ Introduction }
Questions of initial conditions in the universe have
been around for many years, but only now with the promise of detailed
data from the cosmic microwave background can we expect significant
experimental input. The proposal which this work is concerned with is
that of the universe `tunnelling from nothing'. That such a situation 
could be realised  was noticed by Vilenkin \cite{vilenkin82}, where
he pointed out that the compact nature of the de Sitter instanton
could be interpreted as the universe coming from nothing into
a de Sitter phase via the de Sitter instanton. 
In a similar vein Hartle and Hawking \cite{hartle83} attempted 
to describe the initial
conditions of our universe by proposing that the
Lorentzian metric of spacetime be rounded off with a Euclidean
manifold. The two approaches are similar in philosophy but different
in details,  leading to different probability measures for
the initial conditions of the universe\cite{vilenkin98}.

Studying instantons is at some level independent of these 
considerations. What one is looking for is a Euclidean solution
of the field equations which allows a Lorentzian solution to be glued
smoothly onto it, the question of probability measures
comes later.
As such, we start by just trying to find solutions to general
relativity on a manifold with Riemannian metric, a daunting
task as there are an infinite number of degrees of freedom in  the
system. To aid us we impose some symmetry requirements that leave us
with a tractable problem, this is not entirely ad hoc as the analogous
system of a field tunnelling in flat space shows. Coleman {\it et al} 
\cite{coleman78} showed that imposing SO(4) symmetry on a bubble at 
zero temperature was sensible in  that it had the lowest action and 
was therefore more likely to nucleate. 
In fact it was later shown that it was more
than just sensible, because all other solutions to such
a system are singular \cite{garyref}.

The idea of singular solutions has been brought to the fore by
Hawking and Turok \cite{hawking98}; there it was pointed out that 
whereas singular solutions in flat 
space means infinite action, this is not the case in
gravity. Including
gravity means that when the solution starts to blow up the manifold
at least has a chance to close up and make the volume of the instanton
finite, allowing a non divergent action. Whether such solutions are
to be allowed into the path integral has been hotly debated, here we
take the opinion  that we should really let the action decide what
is allowed rather than introducing some notion of singularity by hand
to disallow these instantons. We shall find in fact that this is more
restricting than may be expected. It was asserted in \cite{hawking98} that
generic potentials lead to finite action for solutions with SO(4) symmetry
(although this was later shown to be true only for polynomial and certain
exponential potentials \cite{saffin98}). We will find here that even 
with polynomial potentials the symmetry of Bianchi IX is not always
enough to keep the action regular, unlike imposing SO(4) symmetry.


\section{Einstein equations.}

In this paper we consider the field theory of general relativity
coupled to a scalar field according to the following action,
\begin{eqnarray}
\label{action1}
S_{\rm{E}}&=&\int \eta \left[-\frac{1}{2\kappa}R+\frac{1}{2}(\partial
\phi)^2 +{\cal V}(\phi)\right]+\textnormal{boundary term},
\end{eqnarray}
where the boundary terms are designed to cancel the second derivatives
of  the metric, allowing a sensible meaning to the variational
principle on a boundary \cite{gibbons77}, \cite{barrow89}. The gravitational coupling
is $\kappa=8\pi G$, which we rescale to unity,
and $\eta$ is the volume four-form of the manifold. 
The metric is  taken to be positive definite
and of the form,
\begin{eqnarray}
\label{metric}
\rm{d}s^2&=&\rm{d}\tau^2+a(\tau)^2(\sigma^1)^2+b(\tau)^2(\sigma^2)^2+c(\tau)^2(\sigma^3)^2.
\end{eqnarray}
The $\sigma^i$ are the left invariant one forms of SU(2) and so 
satisfy
\mbox{$\rm{d}\sigma^i=-\frac{1}{2}\epsilon^{ijk}\sigma^j\wedge\sigma^k$}, making
the solution fall into the Bianchi IX class. 
Using Cartan's structure
equations for a torsion free connection it is a nice exercise to find
the curvature two forms,
leading to the following field equations,
\begin{eqnarray}
\label{aeqn}
2\frac{a''}{a}+\frac{a'b'}{ab}+\frac{a'c'}{ac}-\frac{b'c'}{bc}&=&
-\frac{5}{4}\frac{a}{bc}\left(-\frac{a}{bc}+\frac{b}{ac}+\frac{c}{ab}\right)\\
\nonumber &~&+\frac{3}{4}\frac{b}{ac}\left(\frac{a}{bc}-\frac{b}{ac}+\frac{c}{ab}\right)\\
\nonumber &~&+\frac{3}{4}\frac{c}{ab}\left(\frac{a}{bc}+\frac{b}{ac}-\frac{c}{ab}\right)\\
\nonumber &~&-\frac{1}{2}\phi'^2-{\cal V}\\
\label{constraint}
\frac{a'}{a}+\frac{b'}{b}+\frac{c'}{c}&=&
\frac{1}{4}\frac{a}{bc}\left(-\frac{a}{bc}+\frac{b}{ac}+\frac{c}{ab}\right)\\
\nonumber &~&+\frac{1}{4}\frac{b}{ac}\left(\frac{a}{bc}-\frac{b}{ac}+\frac{c}{ab}\right)\\
\nonumber &~&+\frac{1}{4}\frac{c}{ab}\left(\frac{a}{bc}+\frac{b}{ac}-\frac{c}{ab}\right)\\
\nonumber &~&+\frac{1}{2}\phi'^2-{\cal V}\\
\label{phieqn}
\phi''+\frac{\beta'}{\beta}\phi'&=&\frac{\partial {\cal V}}{\partial
\phi}.
\end{eqnarray}
Here \mbox{$\beta(\tau)=a(\tau)b(\tau)c(\tau)$} is the volume of the
surface at constant $\tau$ and $'$ denotes differentiation with
respect to $\tau$. To get the equations of motion for $b(\tau)$ and
$c(\tau$)
we use the $\bf{Z}_3$ symmetry of the equations and just cyclically permute
$a$, $b$ and $c$ in (\ref{aeqn}).

\section{The regular solutions.}

We know that for a potential with a false vacuum there is the 
Coleman-De Luccia solution \cite{coleman80} where we take the functions 
$a(\tau)$, $b(\tau)$ and $c(\tau)$ to be equal, the 
isotropic limit of Bianchi IX.
When we want a regular solution we require it be regular everywhere,
in particular at the origin. In the coordinates we are using one finds
that the Coleman-De Luccia solution has the scale factor behaving as
\mbox{$a(\tau\rightarrow0)\rightarrow \tau/2$} giving the 
instanton a metric which locally is like an isotropic metric on 
{\bf R}$^4$.
This however is not the only way to smoothly end an instanton.
The Coleman-De Luccia solution ends on what Gibbons and Hawking
\cite{gibbons79} called a nut, where all the scale factors vanish
linearly as $\frac{1}{2}\tau$. There is also the case of closing the
instanton off with a bolt, in this case where one of the scale factors
vanishes linearly as $\frac{1}{2}\tau$ the other two become 
equal and constant. One could also consider the situation where the
vanishing scale factor does so with unit proportionality to $\tau$,
this changes the group orbits of the homogeneous submanifolds to
SO(3) rather than SU(2).
Now we consider the regular bolt solution with SU(2) orbits on the 
homogeneous submanifolds, calling the point at \mbox{$\tau=0$} 
the south pole. 
At the bolt then we impose that $c$ varies as $\frac{1}{2}\tau$ and
$a$, $b$ are equal and constant,
with the equations of motion showing
that they remain equal. 

A nice geometrical picture exists for these instantons in terms
of expanding and contracting three spheres. The constant $\tau$ slices
are seen to be homogeneous three spheres, whose anisotropy is 
dictated by the scale factor at that time. The nut-nut solution is 
viewed as an isotropic three sphere expanding from zero size at the
south pole and contracting back to zero at the north pole. Looking
at the bolt-bolt case we see that the ratio of the unequal scale factors
represents the anisotropy of the S$^3$ slices, which diverges at the
poles. This limit of anisotropy is where the three sphere, viewed as
an S$^1$ bundle over S$^2$, has fibres of zero length, 
giving the two sphere limit of a squashed three sphere.
The picture we have then is that at the south pole we have a maximally
squashed  S$^3$ i.e. S$^2$, becoming less squashed as we move off the pole,
finally becoming a two sphere at the north pole.

With the three volume $\beta$ increasing linearly from zero at the
south pole bolt we may infer from the Euclidean Raychaudhuri equation
\cite{jensen89} that for some finite \mbox{$\tau>0$} $\beta$ will
again go to zero, assuming that the scalar potential is positive
semi-definite. The point at which the three volume goes back to 
zero we shall call the north pole.
The manner in which it closes, i.e. singular or not, depends on the
value of the constant that $a(0)\left(=b(0)\right)$ 
is chosen to take, as well as on the
value $\phi(0)$.
We can see that $\phi(0)$ is not fixed by the constraint equation
(\ref{constraint}) by noting that near the bolt 
\mbox{$a\rightarrow \chi+\xi\tau^2$}, with $\chi$ and $\xi$ arbitrary.
The constraint equation then relates $\phi(0)$ to these arbitrary 
constants.
Given that the 3 volume, $\beta$, vanishes and we are looking for
regular $\phi(\tau)$ then (\ref{phieqn}) shows us that 
\mbox{$\phi'=0$} at the north pole, as $\beta'/\beta$ diverges there.

A similar argument to the undershot-overshoot argument of 
\cite{mottola84} suggests that values of $\phi(0)$ and $a(0)$ exist
such that the instanton ends on a bolt at the north pole,
if we have a potential with a false vacuum at 
\mbox{$\phi=0$} and true vacuum for some positive $\phi$.
For a given $a(0)$ we may take $\phi(0)$ to be close to the vacuum
value, in which case $\phi$ will diverge to negative values as we
approach the north pole. If we start further from the vacuum value then
$\phi$ will have turned around by the time the damping
force $\beta'/\beta$ of (\ref{phieqn}) becomes anti damping, driving
$\phi$ to positive infinity. There is thus a value between these which
make the scalar field constant at the north pole. We then find however
that $a(0)$ has not been fixed yet. At the north pole the functions are
now regular, however it
will not in general be a bolt, with $c(\tau\rightarrow\tau_{\rm{N}})$ 
going like $\gamma(\tau-\tau_{\rm{N}})$ and $\gamma$ being different from
a half. We now use the freedom in the value of the constant that
$a(0)$ takes to find the solution with \mbox{$\gamma=\frac{1}{2}$}.

To make these arguments concrete we took the potential
\begin{eqnarray}
{\cal V}(\phi)&=&(\phi^2-1)^2(\phi^2+\alpha),
\end{eqnarray}
which exhibits a false vacuum at the origin for
\mbox{$0<\alpha<\frac{1}{2}$}, we took \mbox{$\alpha=0.1$}.
A regular solution using this potential has been found numerically 
and is displayed below in Fig. 1. We see from the profile of
$\phi(\tau)$ that the scalar field interpolates between the false and
true vacua, as in the Coleman-De Luccia case. However in that case one 
is able to find a section of the instanton which has zero extrinsic
curvature, allowing one to use this surface to attach a Lorentzian
metric to it corresponding to an open universe. 
Here however we do not expect to find such a three-surface
owing to the anisotropy of the homogeneous submanifolds. We can however
find a solution which has an inversion isometry, implying that the
stable surface of this symmetry has zero extrinsic curvature 
\cite{gibbons90}. Such a solution is shown in Fig. 2 where
we see that the profiles are symmetric about the mid point between the
north and south poles. Such an instanton nucleates a closed universe
with anisotropic spatial sections.

\begin{figure}[!htb]
\centerline{\setlength\epsfxsize{80mm}\epsfbox{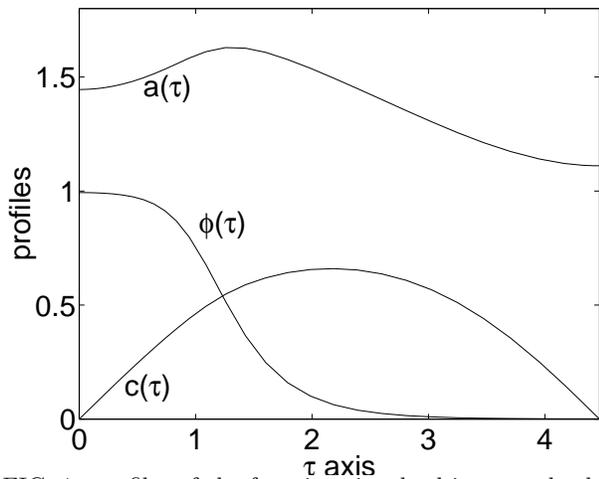}}
 \caption{profiles of the functions involved in a regular bolt
instanton.}
 \label{bolt1}
\end{figure}

\begin{figure}[!htb]
\centerline{\setlength\epsfxsize{80mm}\epsfbox{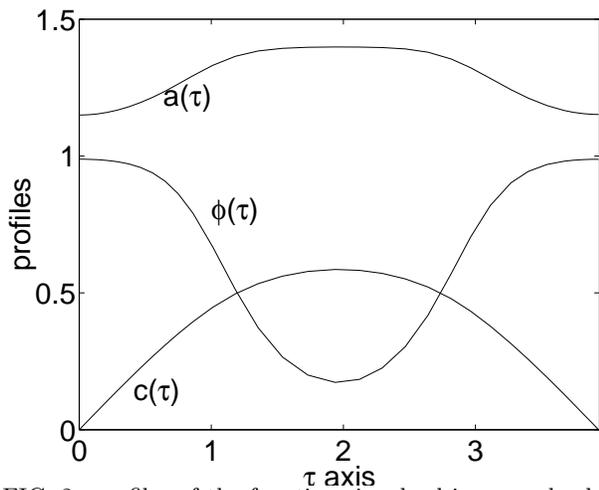}}
 \caption{profiles of the functions involved in a regular bolt
instanton with a parity isometry.}
 \label{bolt2}
\end{figure}

\section{The `conical' solutions}

The suggestion that singular solutions are valid
allows us to consider the case where the metric functions are regular
but the curvature diverges, reminiscent of the curvature blowing
up at the end of a cone. This option was not available in  the 
isotropic case, as the constraint (\ref{constraint}) 
fixed the gradient of the metric
functions. The freedom of the bolt type boundary conditions allows
us to satisfy the constraint with $c(\tau)$ varying linearly with any
gradient and $a(\tau)\left(=b(\tau)\right)$ being an arbitrary
constant. Here we consider such a case, imposing the condition that
the instanton have a parity isometry such that we may attach a real
Lorentzian metric to the stationary surface. On this surface the
scale factors $a$ and $c$ take some constant value and their ratio,
\mbox{$\zeta=a(\tau_{\rm{sym}})/c(\tau_{\rm{sym}})$},
gives a measure of the anisotropy, such that the isotropic case 
has $\zeta$ equal to unity. We then expect that varying $\zeta$ will
allow us to move continuously from the bolt-bolt instanton to
the Coleman-De Luccia solution with parity isometry. This is indeed
the case and is illustrated below in Figs. 3, 4. The upper set of
curves in Fig. 3 are the $a(\tau)$ scale factors as we vary the 
anisotropy measure, $\zeta$, whilst the lower set are the $c(\tau)$.
We see that one is able to interpolate between the two regular
solutions (i.e. the Coleman-De Luccia case with
\mbox{$a(\tau)=c(\tau)$} and the bolt-bolt of Fig. 2)
by using the solutions which have a conical singularity.
To make the comparison clearer in this figure we have scaled all
the solutions to have the same size, whereas in actual fact the
instantons become larger as we approach the isotropic limit. This
information is given in Fig. 4, where we can see how the size,
$\tau_N$,
and the Euclidean action (including boundary terms) varies with
anisotropy. One can see that the action is lowest for the isotropic
case, confirming naive expectations. We also see that the bolt-bolt
instanton has no privileged position, the anisotropy of the bolt-bolt
case is around \mbox{$\zeta=2.4$} at which point the action shows no
special behaviour. We also note here that 
the dotted line corresponds
to the, separately calculated, nun-nut (Coleman-De Luccia) case. 

\begin{figure}[!htb]
\centerline{\setlength\epsfxsize{90mm}\epsfbox{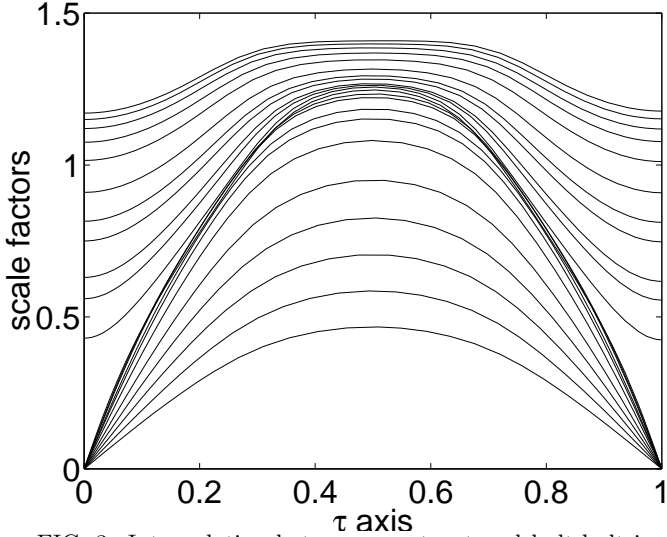}}
 \caption{Interpolating between a nut-nut and bolt-bolt 
instanton. The upper set being the $a$ scale factor and  the 
lower being the $c$.}
 \label{cone}
\end{figure}

\begin{figure}[!htb]
\centerline{\setlength\epsfxsize{90mm}\epsfbox{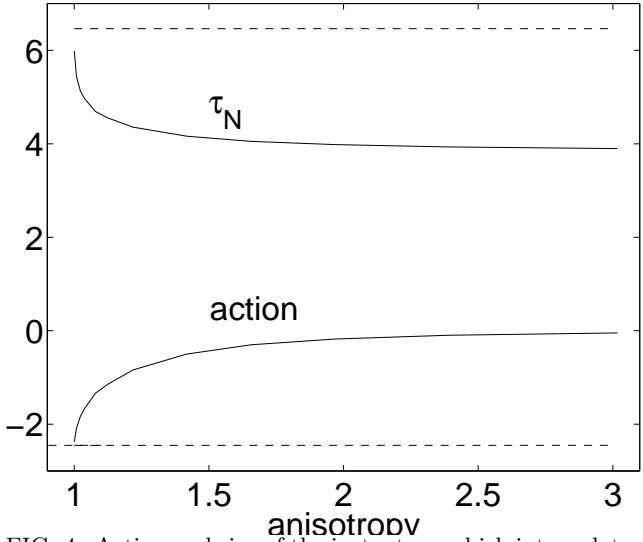}}
 \caption{Action and size of the instantons which interpolate
between the nut-nut and bolt-bolt solution.}
 \label{cone2}
\end{figure}



\section{The singular solutions.}

We now move on to  consider the solutions which possess a singularity
on the manifold
different from the conical type discussed above. 
The type of singularity we are considering here is
the analogue of the
singularity studied in the Hawking Turok paper \cite{hawking98}.
This singularity is different in that the curvature here diverges
as $1/(\tau-\tau_{\rm{N}})$ whilst the conical singularity is more
like a delta function at the poles.

As we are not looking for regular solutions we drop the tunnelling
form of the potential used before and work with a simple quadratic
potential. First we look for solutions which contain a reflection
isometry. This is achieved by setting initial conditions at some
point such that the fields $a$, $b$, $c$ and $\phi$ all have vanishing
gradient at that point, this allows a closed anisotropic universe to
be grafted on to this surface.

We will be interested in finding the Euclidean action for such
instantons
and as such we need to take care of the boundary terms mentioned in
(\ref{action1}). On our solution the singularity must be excised from
the manifold, this means taking a small region around the singularity
and discarding it, naturally introducing a boundary. The boundary term
is then the contribution this boundary makes as the region is made
arbitrarily small.
We find that the Riemann curvature for the metric (\ref{metric}) is,
\begin{eqnarray}
R&=&-2(a''/a+b''/b+c''/c)-2(a'/a+b'/b+c'/c)\\
\nonumber &~&+\frac{1}{4}\frac{a}{bc}\left(-\frac{a}{bc}+\frac{b}{ac}+\frac{c}{ab}\right)\\
\nonumber &~&+\frac{1}{4}\frac{b}{ac}\left(\frac{a}{bc}-\frac{b}{ac}+\frac{c}{ab}\right)\\
\nonumber &~&+\frac{1}{4}\frac{c}{ab}\left(\frac{a}{bc}+\frac{b}{ac}-\frac{c}{ab}\right),
\end{eqnarray}
giving the boundary term of (\ref{action1}) as,
\begin{eqnarray}
\textnormal{boundary
term}&=&\frac{1}{\kappa}\left(\beta'(\tau\rightarrow\tau_S)-\beta'(\tau\rightarrow\tau_N)\right)V_3\\
V_3&=&\int\sigma^1\wedge\sigma^2\wedge\sigma^3.
\end{eqnarray}
In order to calculate with such ill behaved solutions one needs to be
confident of the numerics. To be sure that the correct answers are at least
well modelled we ran a number of simulations with different
integrators. The results presented below are those calculated using a
Bullirsch Stoer method \cite{numrec}, which was tested against a Runge
Kutta routine also from \cite{numrec} and the
integrators found in mathematica and maple. All gave the same results to
fractions of a percent. What one finds in doing this is that this
simple potential does not always give finite action, as it did for
the O(4) invariant solution. 
Following \cite{hawking98} we assume that near the singularity the
form  of the potential becomes irrelevant, a fact backed up by
numerical solutions with polynomial potentials.
The equation for $\phi$ then shows that near the north pole
\mbox{$\phi'(\tau\rightarrow\tau_N)\rightarrow 1/\beta$}.
Taking the constraint equation (\ref{constraint}) assuming
a form of \mbox{$a\rightarrow (\tau_N-\tau)^A$},
\mbox{$b\rightarrow (\tau_N-\tau)^B$},
\mbox{$c\rightarrow (\tau_N-\tau)^C$}, where $A$, $B$ and $C$ are
positive, we then drop the terms which are products of scale factors.
Using the fact that $\phi'$ is varying inversely as the three volume one
finds \mbox{$A+B+C=1$}, showing that dropping the products of
scale factors was consistent. The three volume, $\beta$, therefore goes
to zero linearly, implying that the scalar field diverges
logarithmically, which is slow enough that the action can converge. 

When looking at the action as a function
of initial values for $a$, $b$ and $c$ we found the very striking
result that unless two of the scale factors were equal the boundary
term would diverge. Inspection of  the numerical solution revealed
that for such solutions not all of the scale factors went to zero
at the singularity, this effect is coming about because it is the 
scale factor products, rather than $\phi'$ which dominates in the
singularity. In fact such behaviour can also exist when two  of  the scale
factors are equal, say $a$ and $b$. It is found  that the boundary
term diverges unless \mbox{$c(\tau=0)<a(\tau=0)$}. 
It is interesting to note that this biaxial behaviour where two scale
factors are equal also appeared for the regular solutions, but for
different reasons. 
The regular instantons were required to be biaxial, extending the 
isometry from SU(2) to U(2), so that the instanton could be
closed off at a regular bolt, see also \cite{gibbons90}.
The picture we
then arrive at is given below in Fig. 5, which is a plot of the total
action, when it is finite, for a set of singular instantons. The
figure shows a set of contours indicating that the minimum action occurs
for the isotropic instanton with \mbox{$a(\tau=0)=b(\tau=0)=c(\tau=0)$}.

\begin{figure}[!htb]
\centerline{\setlength\epsfxsize{90mm}\epsfbox{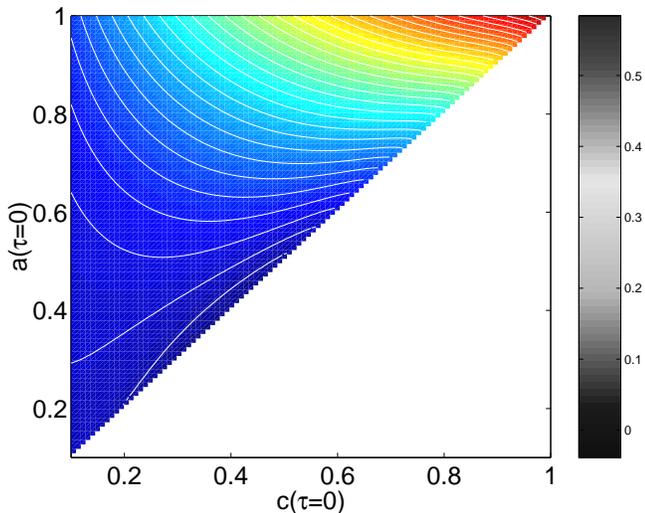}}
 \caption{Total action for the singular instanton with
reflection isometry and \mbox{$b(\tau)=a(\tau)$}, only the
finite action solutions are shown. It is seen that the
minimum action occurs along the line of isotropy.}
 \label{gond1}
\end{figure}


Although this figure only covers the \mbox{$b(\tau)=a(\tau)$} plane,
it appears from the numerical results that it contains all the
relevant information. The $\bf{Z}_3$ permutation symmetry of the 
scale factors mean that this plane is just the same as the 
\mbox{$b(\tau)=c(\tau)$} and \mbox{$a(\tau)=c(\tau)$} planes, with
the action diverging off these planes. We then note that the minima
of the action is to be found along the isotropy axis, that is to say
the metric which has the O(4) isometry and is
the gondola instanton studied in \cite{bousso98}.


\section{Conclusions}
We have discussed a generalisation of the SO(4) symmetric
instantons, keeping the
homogeneity assumption but relinquishing isotropy.
In particular, our solutions were of the Bianchi IX type.
It was found that the singular instantons in this class which allowed a
reflection isometry, leading to the initial conditions 
of a closed universe, 
had a minima of the action when the instanton was isotropic. Moreover,
the probability distribution had a $\bf{Z}_3$ symmetry around the axis
of isotropy so the average instanton would lie on this axis (both
mean and mode). 

An in depth search of types of potential has not been carried out
here, concentrating only on the simplest of potentials. Indeed, the 
results of this study did not reveal any reason to expect other
potentials to deviate from the behaviour found here. This applies
to both the
regular solutions where the potential has a false vacuum and the
singular solutions for generic potentials.
One of the main results is that even for the singular instantons isotropy
would seem to be preferred over anisotropy. The study also sheds light
on the issue of whether or not to allow singular instantons, in that
the action can in fact be a good discriminator of solutions,
which may mean that 
removing solutions by hand is not necessary. 
Namely, we saw that it was the high degree of
symmetry in the Hawking Turok solution that meant the criterion
of finite action would keep
all solutions with polynomial potentials.


\section*{Acknowledgement}

We gratefully acknowledge the help of Gary Gibbons, Stephen Gratton,
Nathan Lepora and Neil Turok for useful discussions. This work was partially
supported by PPARC.





\end{document}